\documentclass[a4paper]{spie}
\usepackage{graphicx,amsmath,amssymb,color,subfigure}
\newcommand{\mc}{\mathcal}
\newcommand{\mbf}{\mathbf}
\title{Theory for entanglement of electrons dressed with circularly
polarized light in Graphene and three-dimensional Topological insulators}
 \author{Andrii Iurov and Godfrey Gumbs 
\skiplinehalf
Hunter college of CUNY, 695 Park Ave, New York, NY, $10065$ \\
The Graduate Center of CUNY; 365 Fifth ave, New York, NY 10016}
\authorinfo{E-mail contact: theorist.physics@gmail.com}
\begin{document}
\maketitle

\begin{abstract}
We have formulated a theory for investigating the conditions
which are required  to achieve entangled states of electrons
on graphene and three-dimensional (3D) topological insulators (TIs).
We consider the quantum entanglement of spins by calculating the
exchange energy. A gap is opened up at the Fermi level
 between the valence and conduction bands in the absence
 of  doping when  graphene as well as 3D  TIs are irradiated
 with circularly-polarized light. This energy band gap is
 dependent on the intensity and frequency  of the applied
 electromagnetic field. The electron-photon coupling also
 gives rise to a unique energy dispersion of the dressed
 states which is different from either graphene or  the
 conventional  two-dimensional electron gas (2DEG).
 In our calculations, we obtained the dynamical
polarization function for imaginary frequencies which is then
employed to determine the exchange energy. The polarization
function is obtained with the use of  both the energy eigenstates and
the overlap of pseudo-spin wave functions. We have concluded that
while  doping has a significant influence on the exchange energy
and consequently  on the entanglement, the gap of the energy dispersions
affects the exchange slightly, which could be used as a good
technique to tune and control  entanglement for quantum information
purposes.

\end{abstract}


\section{INTRODUCTION}
\label{S:I}

Topological insulators (TIs), being a novel quantum state of matter,
were first predicted theoretically in 2006 \cite{Ber} and then observed in
experiment. \cite{kexp} The main feature defining  TIS is the existence of
an insulating gap in the bulk and  topologically protected
conducting states localized on its boundaries \cite{HasanTI, Qi}\,.
Conventional classification of TIs follows from their
geometry. Historically,  two-dimensional (2D)  TIs,
which are also referred to as quantum spin Hall (QSH)
insulators as well as their predecessors quantum Hall states,
represented the first state, which did not fit into Landau-Ginzburg
description of a state of  matter by the symmetry and  order
parameter. The first experimental realization of 2D TIs was given
in \texttt{HgTe}/\texttt{CdTe} quantum wells, where the
non-trivial topological  state can be observed if the thickness
of the sample is more than a certain critical
value ($\backsimeq 6.1\,nm$ for \texttt{HgTe}).

\par
\medskip
From a theoretical point of view, 3D TIs were first predicted
 and later confirmed in \texttt{Bi}$_2$\texttt{Te}$_3$ and \texttt{Sb}$_2$\texttt{Te}$_3$. Similar to the case of 2D TIs,
 they could be characterized by a large insulating gap in the bulk
and the spin-polarized surface states are like graphene, protected
from backscattering. The latter property results from the fact
that in the linear approximation the above mentioned surface states
are described by a Dirac cone. The electrons in 3D TIs are also
represented by  a helical liquid, meaning that the electron spins
 are  perpendicular  to the momentum. Such states cannot appear
 in normal  2D   system with  time-reversal symmetry. Consequently,
  they are also referred to as holographic.

\medskip
\par
It has been demonstrated experimentally\cite{tidetect} that
exposing the surface of a  3D TI to circularly-polarized light
induces  photo-currents. The light polarization could be
used to produce and control photo currents. These currents
represent non-equilibrium properties and are unique for 3D TIs.
These currents may also result in an opportunity
to measure  fundamental physical quantities such as the
Berry phase.  A similar current arising from  topological   states
  could appear as a result of the interaction of 3D TI
surface Dirac electrons with linearly-polarized light.

\medskip
\par
The states with an opened energy gap along with their collective
properties are the central concern of this paper. A geometrical
gap is observed in the case of a finite sized sample in the $x-$direction, say\cite{Finite_Size, main_model}. The advantage of inducing
 a gap with  circularly polarized light is it being tunable,
 i.e., the gap  could be controlled by changing the intensity
 of the radiation. The effects of the gap opening were
considered classically for both graphene \cite{AO,NewOka} and
 TIs.\cite{NewOpticalDora,Lind} Additionally, there   have been
 a number of studies using laser radiation  on single
 layer\cite{r1} and bilayer\cite{Luis1} graphene as well
 as graphene nanoribbons\cite{c1,c2}  reporting the
 gap opening as the result of electron-photon interaction.
 From these considerations, it seems that TIs and gapped graphene
 may have potential applications in devices where spin plays a role.

\par
\medskip

The  rest of the paper is organized as follows. In Sec. \ref {S:1},
we present a brief description  as well as  derive  the
electron-photon dressed states in topological insulators,
which we  previously obtained\,\cite{mmain}  but include
for completeness and to introduce our notation. After that,
we calculate the dynamical polarization function for both
real and imaginary  frequencies. Only the latter may
 be employed  to evaluate the electron exchange energy.
 We compare the response function for interacting electrons
in the random phase approximation (RPA)  at frequencies
on the imaginary axis with that obtained for
 frequencies with a small imaginary part since they are
 used in calculating the correlation energy and collective
 plasma excitations, respectively.   Our results which were
 obtained numerically are in agreement with those
obtained analytically for  plasmons in Dirac-cone
graphene\cite{Wpl,DSpl}, gapped graphene \,\cite{Ppl}
and TIs \cite{Lozpl}, thereby giving credibility  to our
calculations for the exchange energy. It appears that
the effect due to the quadratic term $\backsim \mc{D}k^2$
correction to the linear energy dispersion in wave vector $k$
has little influence on  the overlap structure factor.
This is also demonstrated in Sec. \ref {S:2}. The difference
in the single electron energy dispersions is not negligible
far from the Dirac point. Consequently, this does not affect
the electron polarization function in the long wave limit.
Finally, Sec. \ref {S:3} is devoted to our numerical calculations
 of the electron exchange energy, which has been used  as a
  measure of the electron entanglement in quantum dots\cite{b1,d1}
  and in the troughs formed by  surface acoustic waves.\cite{y1}
  We have demonstrated that the chemical potential (doping)
   $\mu$ has a significant effect on the exchange energy.
    On the other hand, the electron energy gap only
    affects the exchange by a few percent, thus providing
    a novel technique for  sensitively tuning the electron
entanglement, which may receive significant applications in
the field of quantum computing by isolating pairs of dressed
Dirac electrons in quantum dots. We provide some concluding 
remarks in Sec.\  \ref{S:4}.
\par
\medskip


\section{Electron-photon dressed states in graphene and
topological insulators}
\label{S:1}

In this section, we provide a rigorous analytic derivation
and discussion of  dressed states on the surface of a 3D  TI.
 As we show below, the electron interaction with circularly
 polarized photons is the only case for which a complete
 analytic solution may be obtained. Apparently, due to the
specific wave number $k-$dependence of the Hamiltonian,
relatively similar solutions may be obtained
for both graphene and 3D TIs \cite{mmain}\,.
\par
\medskip

Analysis of  dressed stated in graphene \,\cite{Kibis,Kibis2}
showed that electrons in graphene may acquire a gap due
to the interaction with  circularly- polarized photons. The
corresponding wave function is no longer chiral and this leads
to specific tunneling and transport properties\,\cite{mine}\,.

\medskip

The Hamiltonian describing  surface states (at $z=0$) of a
3D TI to  order of ${\cal O}(k^2)$ is given by \,\cite{DSTransp}

\begin{equation}
\mc{H}^{surf}_{3D} =\mc{D}k^2+\mc{A}\sigma\cdot{k} = \left({  \begin{array}{cc}  \mc{D} k^2 & \mc{A}k_-  \\
\mc{A}k_+  &  \mc{D}k^2 \end{array} }\right) \ ,
\label{a1}
\end{equation}
where ${\bf k}=(k_x,\,k_y)$ is the in-plane surface wave vector
 and $k_{\pm}=k_x \pm i k_y$. Here, we take into consideration the
diagonal mass terms $\backsim \mc{D}k^2$ in conjunction with the standard
Dirac cone terms $\mc{A}k_{\pm}$.The energy dispersion associated with
this Hamiltonian is given by $\varepsilon^{\rm surf}_{3D}=\mc{D}k^2+\beta
\mc{A}k$ with $\beta = \pm 1$.

\par
\medskip

Let us now consider electron-photon interaction on the surface
of a 3D TI. We first assume that the surface of the 3D TI is
irradiated by circularly polarized light with its quantized vector
potential given by

\begin{equation}
\label{vpsurf0}
\hat{\mathbf{A}}=\mc{F}_0\left(\mathbf{e}_+\,\hat{a}
+\mathbf{e}_-\,\hat{a}^{\dag}\right)\ ,
\end{equation}
where the left and right circular polarization unit vectors
are denoted by $\mathbf{e}_\pm=(\mathbf{e}_x
\pm i\,\mathbf{e}_y)/\sqrt{2}$,and $\mathbf{e}_x$ ($\mathbf{e}_y$)
is the unit vector in the $x$ ($y$) direction. The amplitude of the
circularly polarized light is related to the photon angular frequency
$\omega_0$ by $\mc{F}_0\sim\sqrt{1/\omega_0}$. Here, we consider a weak field
 (energy $\sim \mc{F}^2_0$)  compared to the photon energy
 $\hbar \omega_0$. Additionally, the total number $N_0$ of
 photons is fixed for the optical mode represented by
 Eq.\,(\ref{vpsurf0}), corresponding to the case with  focused
 light incident on a portion of an optical lattice modeled by
 Floquet theory\,\cite{AO}.

\par
\medskip

We  now turn to the case when the surface of the 3D TI is irradiated
 by circularly polarized light with  vector potential

\begin{equation}
\label{vpsurf}
\hat{\mathbf{A}}={\cal F}_0\left(\texttt{\bf e}_+\hat{a}+\texttt{\bf e}_-\hat{a}^{\dag}\right) \ ,
\end{equation}
where $\texttt{\bf e}_\pm=(\texttt{\bf e}_x \pm i \texttt{\bf e}_y)/\sqrt{2}$,
$\texttt{\bf e}_x$ and $\texttt{\bf e}_y$ are   unit vectors in
the $x$ and $y$ direction, respectively. Consequently, the in-plane
components of the vector potential may be expressed as%

\begin{gather}
\hat{A}_x=\frac{{\cal F}_0}{\sqrt{2}}(\hat{a}+\hat{a}^{\dag})\ ,
\hspace{0.25in} \hat{A}_y=i \frac{{\cal F}_0}
{\sqrt{2}}(\hat{a}-\hat{a}^{\dag}) \ .
\end{gather}
In order to include electron-photon coupling, we make the following
substitution for electron wave vector

\begin{gather}
k_x \longrightarrow k_x +\frac{e\hat{A}_x}{\hbar} = k_x+\frac{e{\cal F}_0}{\sqrt{2}\hbar}(\hat{a}+\hat{a}^{\dag}) \ , \notag \\
k_y \longrightarrow k_y+\frac{e\hat{A}_y}{\hbar} = k_y+i \frac{e{\cal F}_0}{\sqrt{2}\hbar}(\hat{a}-\hat{a}^{\dag}) \ , \notag \\
k_\pm \longrightarrow k_{\pm}+\frac{\sqrt{2}e{\cal F}_0}{\hbar}\,\hat{a}^{\dag}(\hat{a}) \ , \notag \\
k^2=k_+ k_- \longrightarrow k^2+\frac{\sqrt{2}e{\cal F}_0}{\hbar}\left(k_+\hat{a}+k_-\hat{a}^{\dag}\right)
+ \left(\frac{\sqrt{2}e{\cal F}_0}{\hbar}\right)^2
\hat{a}^{\dag}\hat{a} \ .
\end{gather}
In our investigation, we consider high intensity  light with $N_0=\langle\hat{a}^\dag\hat{a}\rangle \gg 1$, and then,
$\hat{a}\hat{a}^{\dag}\sim\hat{a}^{\dag}\hat{a}$ due to
$\hat{a}\hat{a}^{\dag}= \hat{a}^{\dag}\hat{a} + 1$ for bosonic
operators. We adopt this simplification only for the second-order
terms $\sim \left(\sqrt{2}e{\cal F}_0/\hbar\right)^2$
but not for the principal ones containing $\hbar\omega_0$.
With the aid of these substitutions, the Dirac-like contribution
 to the Hamiltonian in Eq.\,\eqref{a1} becomes

\begin{gather}
\mc{H}_{Dirac}= \mc{A}\sigma\cdot{k} = \mc{A}
\left(\sigma_-k_+ + \sigma_ +k_-\right)  \notag \\
\longrightarrow \mc{A} \left({\sigma_- k_+
+\sigma_+ k_-}\right)+\frac{\sqrt{2}e {F}_0}{\hbar}\,\mc{A}
\left(\sigma_-\hat{a}^{\dag}+\sigma_+\hat{a}\right)\ ,
\end{gather}
where $\sigma_{\pm}=(\sigma_x \pm
i \sigma_y)/2$.

Since we are considering the coupling of two quasi-independent
particles (sub-systems) , we are also required to take into account
the photon energy $\hbar\omega_0\,\hat{a}^{\dag}\hat{a}$ in order to have
the appropriate description.

\par
\medskip
This results in the following Hamiltonian:
\begin{gather}
\hat{\mc{H}}=\left(\hbar\omega_0+4\mc{D}\zeta^2\right)
\hat{a}^{\dag}\hat{a}+\mc{D}k^2\mathbb{I}_{[2]}
+2\zeta \mc{D}\left(k_+\hat{a} + k_-\hat{a}^{\dag}\right)
\mathbb{I}_{[2]}   \notag \\
+ \mc{A}  \left(\sigma_+k_-+\sigma_-k_+\right)
+2\zeta\mc{A}\left(\sigma_+\hat{a}
+\sigma_-\hat{a}^{\dag}\right)\ ,
\label{hamil}
\end{gather}
where $\zeta=e{\cal F}_0/(\sqrt{2}\hbar)$. Let us rewrite
the Hamiltonian in Eq.\,\eqref{hamil} in matrix form as

\begin{gather}
\hat{\mc{H}} =\left(\hbar\omega_0+4\mc{D}\zeta^2\right)
\hat{a}^{\dag}
\hat{a}+\underline{\mathbf{1}}+\underline{\mathbf{2}}
+\underline{\mathbf{3}} \notag \\
\equiv\left(\hbar\omega_0+4\mc{D}\zeta^2\right)\hat{a}^{\dag}
\hat{a} +\left({  \begin{array}{cc}  \mc{D} k^2 & \mc{A}k_-  \\
\mc{A}k_+  &  \mc{D}k^2 \end{array} }\right) \notag \\
+2\zeta\mc{D} \left({  \begin{array}{cc} (k_-\hat{a}^{\dag}
+ k_+\hat{a})  & 0  \\
0  & (k_-\hat{a}^{\dag} +k_+\hat{a}) \end{array} }\right)
+2\zeta\mc{A} \left({  \begin{array}{cc}  0 & \hat{a}  \\
\hat{a}^{\dag}  &  0 \end{array} }\right)\ ,
\label{mainhamB}
\end{gather}
where $\underline{\mathbf{1}}\equiv\hat{\mc{H}}^{\rm surf}_{3D}$
denotes the  initial surface Hamiltonian with no electron-photon
interaction, $\underline{\mathbf{3}}$ gives  the principal
effect due to light coupled to electrons (the only non-zero term
at ${\bf k}=0$) and $\underline{\mathbf{2}}$ is the leading term
showing how different are the dressed stated in 3D TIs compared
those in graphene\cite{mine}.

\medskip
\par

We know from Eq.\,\eqref{mainhamB} that the Hamiltonian at
${\bf k}=0$ reduces to the exactly solvable Jayness-Cummings model,
after we neglect the field correction on the order of
${\cal O}(\zeta^2)$. We obtain

\begin{equation}
\label{JC}
\hat{\mc{H}}_{{\bf k}=0}=\hbar\omega_0\,\hat{a}^{\dag}
\hat{a} + 2\zeta \mc{A}\left(\sigma_+ \hat{a} +
\sigma_- \hat{a}^{\dag}\right) \ .
\end{equation}

Now, we are in a position to expand the sought after wave functions
over the eigenstates of Jayness-Cummings model Eq.\,\eqref{JC} as

\begin{gather}
\vert \Psi^0_{\uparrow, N_0} \rangle = \mu_{\uparrow, N_0} \vert \uparrow, N_0 \rangle +\nu_{\uparrow, N_0} \vert \downarrow, N_0+1 \rangle \ , \notag \\
\vert \Psi^0_{\downarrow, N_0} \rangle = \mu_{\downarrow, N_0} \vert \downarrow, N_0 \rangle - \nu_{\downarrow, N_0} \vert \uparrow, N_0-1 \rangle  \ .
\label{basisdef}
\end{gather}
and obtain the energy eigenvalues as

\begin{gather}\
\frac{\varepsilon^0_{\pm}}{\hbar\omega_0}= N_0 \pm \frac{1}{2} \mp \frac{1}{2}\,\sqrt{1+\frac{\alpha^2}{N_0}\left(N_0 + \frac{1}{2} \pm \frac{1}{2}\right)} \notag \\
\simeq N_0 \pm \frac{1}{2} \mp
\left({\frac{1}{2} +\frac{1}{4}\,\alpha^2}\right)
= N_0\mp \frac{\alpha^2}{4}\ ,
\label{JCdisp}
\end{gather}
where $\alpha^2=2\zeta\mc{A}N_0/(\hbar\omega_0)$ with $N_0\gg 1$.
The energy gap at ${\bf k}=0$ has been calculated as
$\Delta^0\equiv\varepsilon^0_- - \varepsilon^0_+
\approx (\alpha^2/2)\,\hbar\omega_0$. We note that there is
no difference between graphene and the surface states of 3D TI
at ${\bf k}=0$. We assume $\alpha \ll 1 $ and $N_0 \gg 1$,
corresponding to a classically large number of lase photons and
weak light coupling to electrons as a perturbation to the
electron energy. Therefore, we conclude that the effect of electron-photon
interaction is quite similar to graphene as far as one photon
number $N_0$ is concerned. The main difference being that
the energy gap in the 3D TI is of  order  ${\cal O}(\zeta^2)$,
which may be neglected for low intensity light. Consequently,
the energy dispersion relation becomes

\begin{equation}
\varepsilon_{\beta}(k,\,\Delta_0) = N_0\,\hbar \omega_0 + \mc{D} k^2 + \beta \sqrt{\left[\Delta_0 + {\cal O}(\zeta^2)\right]^2
 + \left({\mc{A}k}\right)^2} \ ,
\end{equation}
where $\beta=\pm 1$ and $\Delta_0$ is the photon-induced energy
gap as in graphene.

\begin{figure}
\centering
\includegraphics[width=0.5\textwidth]{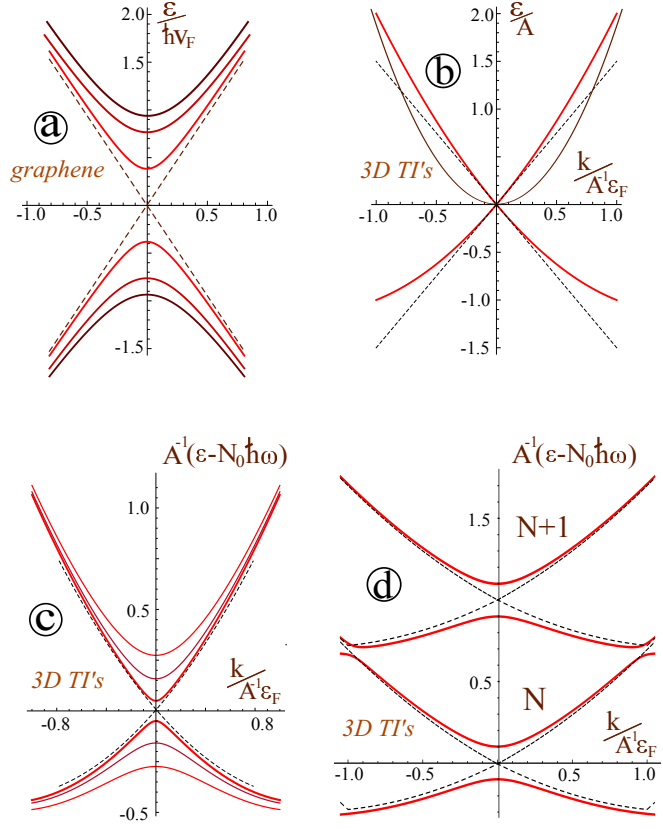}
\caption{(Color online). Energy dispersion relations for
electron dressed states. Plot (a) depicts  the energy
spectrum for graphene under various intensities of circularly
polarized light illumination. The dashed line are for standard
Dirac cone with no electron-photon interaction. Plot (b)
 shows the energy dispersion (red solid curves) for the effective
 surface model of 3D TI without light-electron interaction
 ($\alpha = 0$, $\mc{D} k^2_F/(\hbar \omega_0) =
 \mc{A} k_F/(\hbar \omega_0)= 0.2]$; (c) for single-mode dressed states [$\alpha=0.05$ (inner), 0.07 (middle), 0.1 (outer)]
and (d) for two-mode dressed states. The dashed curves in (c)
 and (d) represent the asymptotic behaviors obtained by setting
 $\alpha=0$.}
\label{f1}
\end{figure}

The ${\cal O}(\zeta^2)$ correction makes an important physical
difference  between the dressed states in 3D TIs and graphene.
However, the  correction is small numerically and may be
neglected in most calculations.The  expression
for the energy spectrum may be  expressed as

\begin{equation}
\varepsilon^{\rm surf}_{3D}(\Delta) = N_0\,\hbar \omega_0
+ \mc{D} k^2 + \beta \sqrt{\Delta^2
+ \left({\mc{A}k}\right)^2} \ ,
\label{singlemodedispersion}
\end{equation}
with $\beta=\pm 1$ and the induced energy gap defined by

\begin{equation}
\Delta = \sqrt{\mc{W}_0^2+(\hbar \omega_0)^2} - \hbar \omega_0
\sim \hbar \omega_0\left(\frac{\alpha^2}{2}\right)\ ,
\end{equation}
where $\alpha=\mc{W}_0/(\hbar\omega_0)$ and $\mc{W}_0$ is the
electron-photon interaction energy. For the upper subband
with $\beta=1$ in Eq.\,\eqref{singlemodedispersion}, the
energy gap is related  to the effective mass around ${\bf k}=0$
through $2 m^\ast_\Delta=\hbar^2/\left[{\mc{A}^2
/ (2 \Delta) + \mc{D}}\right]$, where the photon dressing
decreases the effective mass. This is in contrast with
single-layer graphene, where electron-photon interaction
leads to an   effective mass. A similar phenomenon
 on the  effective mass reduction is also found in bilayer
 graphene under the influence of  circularly polarized light.
We will consider the biggest possible value for $\mathcal{W}_0$
to maximize the light-coupling effect, although the condition
 $\mathcal{W}_0 < \hbar \omega_0$ must be maintained to ensure
 the valid approximations made in this paper. Here, as an example,
 we will just use the leading-order Dirac cone term ${\cal A}\,\sigma
 \cdot{\bf k}$ to estimate the light-induced energy gap. The small
 correction from the parabolic ${\cal D}$ term can be neglected for
 not very large $k$ values.

\par
\medskip

The dressed state wave function corresponding to
Eq.\,(\ref{singlemodedispersion}) is given by

\begin{gather}
\label{wf}
\Phi^{\bf k}_{\rm e-ph}(x,\,y)=\frac{1}{\sqrt{1+\gamma^2(\beta)}}
\left({ \begin{array}{c}    1 \\
\gamma(\beta)\texttt{e}^{i \phi}  \end{array}  }\right)
\texttt{e}^{ik_x x + ik_y y}\ ,
\end{gather}
where $\gamma(\beta)=\mc{A}k/[\Delta+\beta
\sqrt{\Delta^2+(\mc{A}k)^2}]$ and $\phi=\tan^{-1}(k_y/k_x)$.

In conclusion, for a single mode dressed state, the effect
due to the electron-photon interaction is quite similar to graphene.
The difference is  that the energy gap varies as  $\alpha^2$.
 However, this dependence becomes negligible under low -intensity light
illumination. The energy dispersion relations for single  and
double-mode dressed states of 3D TIs are presented in Fig.\,\ref{f1}.
Comparing Figs.\,\ref{f1} (c) and (d), we find that an energy
gap is opened at $k = 0$ due to photon dressing, and the Dirac
cone is well maintained except for large k values. In contrast,
for double-mode dressed states in Fig.\,\ref{f1}(d), additional
mini-gaps appear at the Fermi edge and new saddle points are formed
at $k = 0$ due to strong coupling between dressed states with
different pseudo-spins. These new mini-gaps and the saddle
points prove to have a significant effect on electron transport
and many-body properties. There exists a direct relationship
 between the circularly-polarized \textit{light intensity} and
 the magnitude of energy gap $\Delta$ which is attributed to the
electron-photon interaction. Here, as an example, we will
just use the leading-order Dirac cone term
${\cal A}\,\sigma \cdot{\bf k}$ to estimate the light-induced
energy gap.  The small correction from the parabolic
${\cal D}$ term can be neglected
for not very large $k$ values.

\par
\medskip
For a circular-polarized $CO_2$ laser beam with power
$P \backsim 10^2\;W$, the wavelength
$\lambda \backsim 10^{-5}\;m$ and the  beam size on the order
of $\backsim \lambda$, from the field energy density $w \backsim P/(\lambda^2c)$
we find the electric field amplitude $F_0 \backsim \sqrt{P/(\epsilon_0\lambda^2c)}
= 10^5 V/cm$. This leads to the light-coupling energy $W_0 \backsim 10^{-20}\:J$
and the light-induced energy gap $\Delta\backsim 0.01 - 0.1 \; eV$ for $Bi_2Se_3$
with $\mathcal{A}=10^{-27}\;J\;cm$.


\section{Dynamical polarization and plasmons}
\label{S:2}

\begin{figure}
\centering
\includegraphics[width=0.6\textwidth]{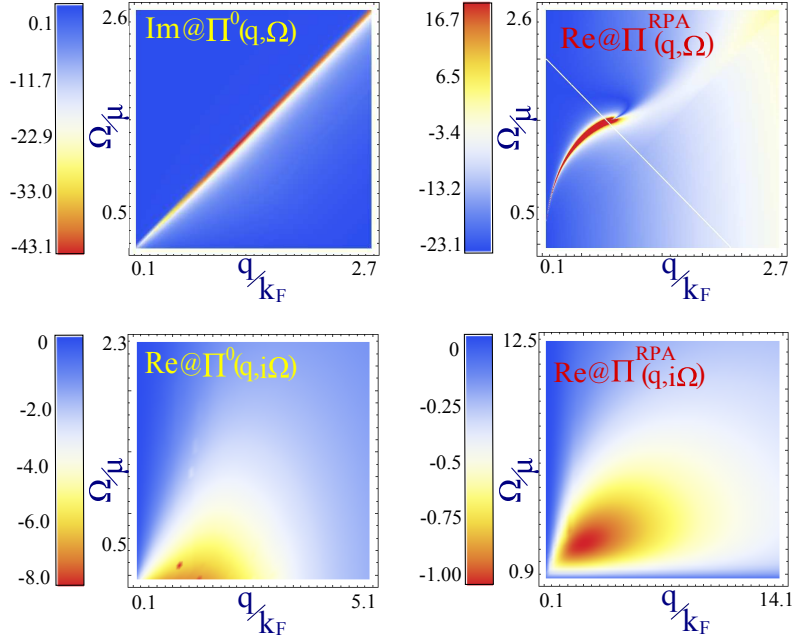}
\caption{(Color online). Density plots for zero band gap 
of the polarization function
for real and imaginary frequencies. The upper left panel
presents the imaginary part of the non-interacting  polarization
function at real frequency $\Omega$. This plot  shows the
region  where the plasmons are undamped, i.e., Im$[\Pi(q,\Omega)]=0$.
The upper right plot is the real part of the interacting
polarization $\Pi^{RPA}(q,\Omega)$, which shows the  plasmon
dispersion $ \backsim \sqrt{q}$. The lower panels are the
two corresponding plots for  imaginary frequencies $i \Omega$.
The lower left plot is the real part of the non-interacting
 polarizability; its imaginary part  is identically zero.
 The lower right plot is the RPA polarization function for
 imaginary frequency. The plots described a gapless topological insulator.}
\label{f2}
\end{figure}

For us to calculate the electron exchange energy, we need
to obtain the non-interacting  polarizability $\Pi^{0}(q, i \Omega)$
for imaginary  frequency $i \Omega$ and wave number $q$.
Let us consider the case of chemical potential $\mu>0$ such
that the occupied electron states partially fill  a finite
part of the conduction  band up to a certain   value $\mu$.
Then,

\begin{equation}
\label{mainpol}
\Pi^0(q,i\Omega)=\frac{1}{\pi^2} \sum_{s,s'=\pm 1} \int_{\Lambda} d^2 k \; \mc{F}(\mbf{k},\mbf{q}) \;
\frac{N_F[\mc{E}_s(k)]-N_F[\mc{E}_s'(k+q)]}{\mc{E}_s(k)-\mc{E}_s'(k+q)
+i\hbar   (\Omega+\gamma)}\ ,
\end{equation}
where $\gamma$ is real. The polarization function, obtained
analytically in \cite{Wpl} for real frequencies is not suitable for
our calculations. There,   the only  imaginary
term  is infinitesimal $i\gamma$.
The structure factor $\mc{F}(k,q)$ is given as:

\begin{equation}
\label{prefactor}
\mc{F}(\mbf{k},\mbf{q})=\vert \langle \Psi_s (k)
\vert \Psi_s'(k+q) \rangle \vert^2 \ .
\end{equation}
According to Eq.\ref{wf}, the electron dressed state wave functions may
 be expressed  as

\begin{equation}
\Psi^{T}_{s}(k)= \{ C_1(k);\;C_2(k) \texttt{e}^{i \theta_k} \}
\end{equation}
and, correspondingly,

\begin{equation}
\Psi^{T}_{s'}(k+q)= \{ C_1(k+q);\;C_2(k+q) \texttt{e}^{i \theta_{k+q}} \}
\end{equation}
so that the equation for the structure factor [\ref{prefactor}] becomes

\begin{equation}
\mc{F}(\mbf{k},\mbf{q})=C_{1}(k)C_{1}(k+q) + s\;s'\;\texttt{e}^{i \delta_{\theta}}
\end{equation}
with $\delta_{\theta}=\theta_k-\theta_{k+q}$. The prefactor, which
is the squared absolute value
of the wave functions overlap, now reads:

\begin{equation}
\mc{F}(\mbf{k},\mbf{q})=\sum_{\nu=1}^{2}
\left( C_{\nu}(k)C_{\nu}(k+q) \right)^2 + 2
\prod_{\xi=1}^{2} C_{\xi}(k) C_{\xi}(k+q) \; s s' \sin \delta_\theta\ .
\end{equation}
According to Eq. (\ref{wf}),

\begin{equation}
C_1 (k) = \frac{\mc{A}k}{\sqrt{
2 \left( \Delta^2 + \left( A k \right)^2 \right)
 - 2 \Delta \sqrt{\Delta^2 + \left( A k \right)^2}
}}
\end{equation}
and, correspondingly,

\begin{equation}
C_2 (k) = \frac{
\sqrt{\Delta^2 + \left( A k \right)^2} - \Delta
}{\sqrt{
2 \left( \Delta^2 + \left( A k \right)^2 \right)
 - 2 \Delta \sqrt{\Delta^2 + \left( A k \right)^2} \, .
}}
\end{equation}
As one can see, when there is  no energy gap $\Delta \to  0$,

\begin{equation}
 C_1 = C_2 = \frac{1}{\sqrt{2}}
\end{equation}
so that the coefficients no longer depend on the wave vector $k$, 
and the structure factor becomes
 
\begin{equation}
\mc{F}_{\Delta = 0}(\mbf{k},\mbf{q}) =  
\frac{1}{2} \left(1 + s s' \frac{k + q \cos
\phi}{\vert \mbf{k}+\mbf{q} \vert } \right) \, .
\end{equation}
It is instructive to compare the obtained numerical results 
with the  case of real frequency $\Omega$.   Both cases are 
presented in Fig.\ref{f2}. One may conclude that the non-interacting 
polarization $\Pi^{0}(q,i \Omega)$ is non-zero only in a 
small region of   $q - \Omega$ space. On the other hand,
 for real frequency the peak of the polarization function 
 (both of its real and imaginary parts) is located along the 
 main diagonal $\Omega \sim  q$. The renormalized RPA polarization
 function is finite in a certain region and its maximum
no longer represents a line of plasmons.

\medskip
\par
In the region, where undamped plasmons exist $(\Omega > \mc{A} q $
 and $\Omega < \mu - \mc{A} q)$ where $\mc{A}$ is a constant,
  the real frequency polarization for interacting electrons in the RPA may be expressed as
  
\begin{equation}
\Pi(\Omega,q) = \frac{2 e^2 \mu}{\epsilon_0 \mc{A}^2} - \frac{ e^2 \left( \mc{A} k \right)^2 }{4 \epsilon_0 \sqrt{\left( \mc{A} k \right)^2 + \left( \hbar \omega \right)^2 }}
\left({
\mc{O}_+\sqrt{1-\mc{O}_+^2}+\mc{O}_-\sqrt{1-\mc{O}_-^2}+ i \cosh^{-1}(\mc{O}_+)- i \cosh^{-1}(\mc{O}_-)
}\right)
\end{equation}
where $\mc{O}_{\pm}\equiv \frac{2 \mu \pm \hbar \Omega}{\mc{A} q}$.
The resulting plasmon dispersion is determined by the following identity:
\begin{equation}
\frac{1}{q}\Pi(\Omega_{pl},q)+\frac{2 \epsilon_0}{e^2}=0 \, ,
\end{equation}
in the long wavelength approximation resulting in \cite{Ppl}

\begin{equation}
\Omega_{pl}(q)=\left({2 e^2 \epsilon_0 \mu \mc{A} \left({1-\frac{\Delta^2}{\mu^2}}\right) q}\right)^{1/2}
\end{equation}
for $\Delta \to  0$ we obtained the plasmon dispersion for Dirac
cone (graphene)

\begin{equation}
\Omega_{pl}(q)=(2 e^2 \epsilon_0 \mu \mc{A} q)^{1/2} \backsim 
\sqrt{q} \ .
\end{equation}
Clearly, the $\backsim \sqrt{q}$ dependence is similar to what 
we observe in 2DEG. It appears that the plasmon dispersion
in 3D TIs also follow the $\backsim \sqrt{q}$ law.\cite{Lozpl}

\medskip

\par

\section{Exchange energy, theory of entanglement}
\label{S:3}

In this section, we calculate the exchange energy\cite{YYY} as well as 
discuss the entanglement properties. The basic idea is that the 
efficiency of the electron entanglement is directly related 
to the exchange energy $\mc{E}_{ex}$. Therefore,
calculating it is now  our  goal since we may then confine electrons
to quantum dots\cite{XXX} for the purpose of using entangled spins in 
applications such as quantum computing and sensors.

\medskip
\par
The general formula for the electron exchange energy in 3D 
is\cite{Harris1,Griffin1}

\begin{equation}
\mc{E}_{ex}=-\frac{A}{2} \int d^3\mbf{r} \int d^3\mbf{r'}\
 V(\vert \mbf{r}-\mbf{r'} \vert) \left({
\frac{1}{\pi} \int_{0}^{\infty} d\omega \Pi^{0}(\mbf{r},\mbf{r'},i \Omega)+ n(\mbf{r})\delta(\mbf{r}-\mbf{r'})
}\right)
\label{ex1}
\end{equation}
where $A$ is a normalization area, $V(\vert \mbf{r}-\mbf{r'} \vert) $
is the Coulomb interaction and $n(\mbf{r})$ is electron number density.
After a 2D Fourier transformation along 
the $x-$ and $y-$ axes, we obtain

\begin{equation}
\mc{E}_{ex}=-\frac{A}{2} \int d\mbf{k}_{\parallel} \int_{0}^{\infty} dz \int_{0}^{\infty} dz' V(\mbf{k}_{\parallel}, z-z') \left[{
\frac{1}{\pi}  \int_{0}^{\infty} d\Omega \Pi^{0}(\mbf{k}_{\parallel},z,z';i \Omega ) + n_{av}(z)\delta (z-z')
}\right]
\end{equation}
with $n_{av}(z)=(1/A) \int dx\, dy \, \ n(x,y,z)$. For a purely 2D system, 
Eq.\ref{ex1} becomes

\begin{equation}
\mc{E}_{ex}^{surf}=-\frac{A}{2} \int_{-\infty}^\infty dq_x 
\int_{-\infty}^\infty  dq_y\  V(\mbf{q}) \left[
\frac{1}{\pi} \int_0^{\infty}d\Omega\  \Pi^{0}(\mbf{q},i\Omega)+n_{av}
\right]
\end{equation}
here $n_{av}=(1/A) \int dx \, dy \,\  n(x,y) $ is the 2D electron density.The $d^2 q$ integral diverges only for $n_{av} = 0$, which we will accept for the rest
of our calculations.
\par
\medskip

\begin{figure}
\centering
\includegraphics[width=0.45\textwidth]{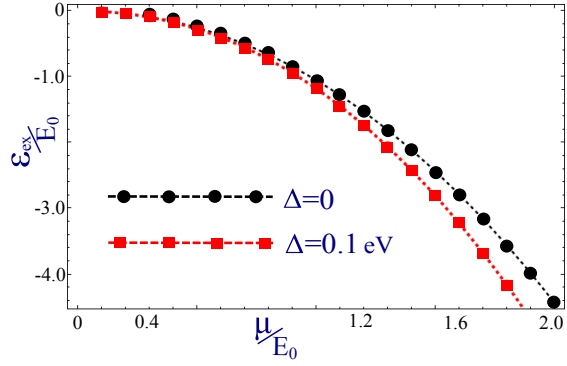}
\caption{(Color online). Electron exchange energy as a 
function of  chemical potential. The \textit{black line} corresponds
to the case of gapless energy dispersion and the \textit{red line} 
is for  energy gap $\Delta=0.9 \mu$. $E_0 \backsimeq 1 \; eV$ is an energy constant, determined from the
integration cut off.}
\label{f3}
\end{figure}
The correlation energy for both graphene\cite{mcd}  and TI  may
 be obtained with the use of  the  PRA polarization for
  imaginary frequencies. The correlation leads to 
  the spin and charge susceptibilities being 
suppressed for   chiral Dirac cone in both graphene and 3D TIs
\medskip
\par
\begin{figure}
\centering
\includegraphics[width=0.43\textwidth]{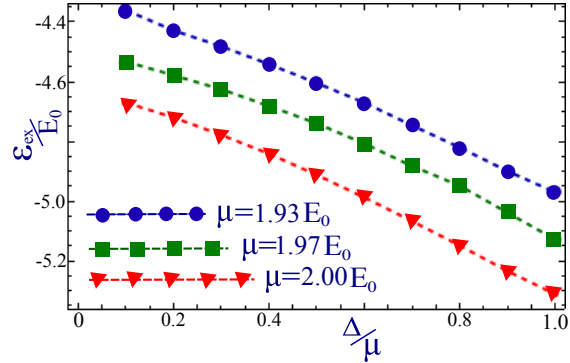}
\caption{(Color online).   Exchange energy is plotted 
as a function of  the energy gap, for chosen 
 chemical potential. The gap for the electron energy 
 dispersion is   scaled with the chemical potential.
Each line corresponds to certain level of electron 
doping, and the and the difference is a few percent.   
This shows that the chemical potential has much 
stronger influence on the  exchange compared to energy gap.}
\label{f4}
\end{figure}
The obtained variation of the exchange energy is presented 
in Figs. [\ref{f3}
It has been argued that the entanglement is determined by the electron exchange energy \cite{b1,d1}. The
entanglement of the electron spins in the channels driven by the surface acoustic waves was considered in \cite{y1} \, .

\section{Concluding Remarks}
\label{S:4}

\medskip
\par
In summary, we have calculated the exchange energy for 
 electron-photon dressed effective surface states for 
 3D TIs. The dressed states, which appear as a result 
 of the electron interaction with circularly polarized 
 light, lead to the existence of a finite gap in the energy 
 dispersion and breaking of the chiral symmetry of the 
 corresponding wave function. The energy gap, which may
  be as large as  $0.1$ eV for circularly-polarized 
  laser beam of $10^2$ W power. This may exceeds the 
  geometrical energy gap which appears if a 3D TI
   sample has a finite width.

\medskip
\par
The  circular polarization of the imposed light allows 
 complete analytic solution for  both  graphene
and   TIs due to the fact that for $k=0$ each Hamiltonian 
coincides with that for  the Jayness-Cummings model and
the corresponding eigenstates are used as a basis of the 
expansion of the  wave functions.

\par
The obtained dressed states in  TIs acquire an energy gap 
like  graphene. However, unlike graphene, the
electron-photon interaction has significant influence 
on the valence (lower) subband. In the case of a higher
interaction amplitude, the hole-subband becomes nearly 
dispersionless. As far as gap is concerned, its value 
is large compared to the  case of the electron dressed 
states in grpahene, but the difference is not important 
or of significant value.

\medskip
\par
We have calculated numerically the non-interacting  polarization 
function with both real and imaginary frequencies. The latter
quantity $\Pi(q,i \Omega)$ enters into the expression for 
the exchange energy and the real $\Omega$ polarization 
together with the  RPA polarization determine the plasmon 
dispersion as well as the region where undamped plasmons could exist. Our calculations
of the real-$\Omega$ polarization completely agree with the earlier obtained results for the plasmonics \cite{Wpl,Ppl}.

\par
It has been argued that the electron exchange energy   
 has a direct  and strong influence on the  entanglement.
This  has been considered for both quantum dots and 
channels, originating from surface acoustic waves.
We concluded that the doping value (chemical potential) 
has a strong influence on the exchange energy and, as a 
consequence, on the electron entanglement. We have 
also found that the energy gap,  due to the electron-photon 
interaction leads to an increase in the  magnitude
 of the electron exchange energy. However, this 
 dependence is much  weaker compared  to the $\mu-$dependence.
As a result, we propose a technique to tune and control 
the electron entanglement by both the chemical potential 
and the electron-photon interaction, so that the second 
mechanism may be used for a small  tuning which we 
believe has a strong potential for device applications.
\par

\section*{ACKNOWLEDGEMENTS}
This research was supported by  contract $\#$ FA $9453-11-01-0263$
of AFRL. The authors also acknowledge considerable contribution and helpful
discussions with \textit{Liubov Zhemchuzhna}.

\bibliographystyle{spiebib}
\bibliography{SBib}
\end{document}